\documentclass[11pt]{article}
\usepackage{amsmath,amssymb,epsfig,cite}
\advance\hoffset by -1.1truecm
\advance\voffset by -1.0truecm
\newcommand{\be}{\begin{equation}}
\newcommand{\ee}{\end{equation}}
\newcommand{\bea}{\begin{eqnarray}}
\newcommand{\eea}{\end{eqnarray}}

\numberwithin{equation}{section}

\newcounter{appendice}

\begin{document}

\title{\begin{flushright}
 \small SU-4252-835\\IISc/CHEP/5/06
\end{flushright}
\vspace{0.25cm} QED on the Groenewold-Moyal Plane}
\author{A. P. Balachandran$^a$\footnote{bal@phy.syr.edu} ,
A. Pinzul$^a$\footnote{apinzul@phy.syr.edu} ,
  B. A. Qureshi$^a$\footnote{bqureshi@phy.syr.edu}
  and S. Vaidya$^b$\footnote{vaidya@cts.iisc.ernet.in}\\ \\
$^a$\begin{small}Department of Physics, Syracuse University, Syracuse NY,
13244-1130, USA. \end{small} \\
$^b$\begin{small}Centre for High Energy Physics, Indian
Institute of Science,
Bangalore, 560012, India.
\end{small}}
\date{\empty}

\maketitle
\begin{abstract}

We investigate a version of noncommutative QED where the interaction
term, although natural, breaks the spin-statistics connection. We
calculate $e^- + e^- \rightarrow e^- + e^-$ and $\gamma + e^-
\rightarrow \gamma + e^-$ cross-sections in the tree approximation and
explicitly display their dependence on $\theta^{\mu \nu}$. Remarkably
the zero of the elastic $e^- + e^- \rightarrow e^- + e^-$
cross-section at $90^\circ$ in the center-of-mass system, which is due
to Pauli principle, is shifted away as a function of $\theta^{\mu
\nu}$ and energy. 

\end{abstract}

\section{Introduction}

Although the defining relations 
\begin{equation}\label{commutator}
[x^\mu ,x^\nu ]= i\theta^{\mu\nu}\ ,\ \ \theta^{\mu\nu} = const
\end{equation}
of the GM or noncommutative plane are not invariant under naive
Lorentz transformations, they are so under the so-called twisted
action of the Poincar\'e group \cite{cknt,Wess:2003da}. This
observation makes it possible for one to study tensorial objects like
scalars, spinors, vector fields, and so on, and in quantum theory,
regain the use of Wigner's classification of particles in
$3+1$-dimensions according to the unitary irreducible representations
(UIRs) of the Poincar\'e group. Indeed, quantum fields that respect
this twisted action (``twisted quantum fields'') can be used to study
quantum field theories (QFTs) on the noncommutative plane
\cite{bmpv}. Twisted quantum fields deviate from the usual
spin-statistics connection at high energy, and it is precisely this
deviation that provides simple interacting theories with immaculate
high energy behaviour: there is no UV-IR mixing \cite{bpq}.

New gauge theories on the GM plane that are compatible with twisted
Poincar\'e symmetry can now be constructed
\cite{Balachandran:2007kv}. This construction works for arbitrary
gauge groups (and not just $U(N)$), with matter fields in any
representation of the gauge group. As a result, it is possible to do
realistic model-building for extensions of the Standard Model, and
look for new phenomenological signals. This formulation takes
advantage of the fact that there is a representation of the
commutative algebra ${\cal A}_0 ({\mathbb R}^N)$ on ${\cal A}_\theta
({\mathbb R}^N)$ where ${\cal A}_\theta ({\mathbb R}^N)$ is the
algebra generated by $x^\mu$'s subject to the relation
(\ref{commutator}). We can thus take the group of gauge
transformations ${\cal G}$ to be based in ${\cal A}_0 ({\mathbb
R}^N)$. The covariant derivative of matter fields respects the module
property and Poincar\'e covariance, as it should.

Since the gauge group is based in ${\cal A}_0 ({\mathbb R}^N)$, the
quantum Hamiltonian for pure gauge theory is the same as the
corresponding commutative one. It is the matter-gauge Hamiltonian that
is different, and remarkably, scattering processes that involve these
two types of interactions break Lorentz invariance
\cite{Balachandran:2007kv}, and carry signals for deviations from the
spin-statistics connection. However, this breakdown of Lorentz
invariance is a very controlled one, in the sense that it can be
described as a manifest quasi-Hopf Lorentz symmetry of the full
Hamiltonian of the theory \cite{Balachandran:2009an}.

Such effects are expected to persist even in models which display
spontaneous symmetry breaking of a non-Abelian gauge group, and in
particular the Standard Model. To illustrate the kind of
phenomenological consequences one might see, we will examine a model
of quantum electrodynamics (QED) on the GM plane, where the
gauge-invariant interaction term, although natural, does not respect
spin-statistics. [This model differs from the ones we have formulated elsewhere. 
We will indicate the differences later.] We explicitly establish that $\theta^{\mu \nu}$ has
physical effects in the interaction of gauge and matter fields by
calculating $e^- + e^- \rightarrow e^- + e^-$ and $\gamma + e^-
\rightarrow \gamma +e^-$ cross-sections and showing their dependence
on $\theta^{\mu \nu}$.

In $e^- -e^-$ scattering (with all electron spins identical) in the
centre-of-mass frame, the amplitude (and hence the cross-section)
vanish for $90^\circ$ scattering angle for $\theta^{\mu \nu}=0$. This
zero is due to Pauli principle. It is moved from $90^\circ$ when
noncommutativity is introduced (There is dependence of $O(\theta)$ on
energy and momentum variables.).  Such a movement of zero shows
violation of Pauli principle.

This article is organized as follows. After a brief review of twisted symmetry in 
section 2, we will specialize in section 3 to the case of QED, where the interaction 
term does not respect spin-statistics
connection. Implications to M\"oller and Compton scattering will be
elaborated, and the we will show that the scattering cross-section
depends on $\theta_{\mu \nu}$.

A final and important fact is brought out in section 4. The
perturbative $S$-matrix is not Lorentz invariant despite all our
elaborate efforts to preserve it. (However it {\it is} unitary,
consistently with \cite{bdfp}.) It is not difficult to understand the
origin of such non-covariance.  The density $H_I$ of the interaction
Hamiltonian is not a local field in the sense that
\begin{equation}
[H_I (x), H_I (y)] \neq 0, \quad x \sim y \label{Bcausal}
\end{equation}
where $x \sim y$ means that $x$ and $y$ are space-like separated.
But $S$ involves time-ordered products of $H_I$ and the equality
sign in (\ref{Bcausal}) is needed for Lorentz invariance. This
condition on $H_I$, known as Bogoliubov causality \cite{bog}, has
been reviewed and refined by Weinberg
\cite{weinberg1,weinbergbook1}. The nonperturbative LSZ formalism
\cite{weinbergbook1} also leads to the time-ordered product of
relatively non-local fields and is not compatible with Lorentz
invariance. Such a breakdown of Lorentz invariance is very
controlled.  For this reason, such Lorentz non-invariance may
provide unique signals for non-commutative spacetimes, a point
which has been elaborated in \cite{Balachandran:2007yf}.

\section{Review of Twisted Quantum Fields}

The algebra ${\cal A}_\theta ({\mathbb R}^N)$ consists of smooth
functions on ${\mathbb R}^N$ with the multiplication map
\begin{eqnarray}
m_\theta: {\cal A}_\theta ({\mathbb R}^N) \otimes {\cal A}_\theta
({\mathbb R}^N) &\rightarrow& {\cal A}_\theta ({\mathbb R}^N)
\nonumber \\
\alpha \otimes \beta &\rightarrow& \alpha \;e^{\frac{i}{2}
  \overleftarrow{\partial}_\mu \theta^{\mu \nu}
  \overrightarrow{\partial}_\nu} \;\beta \equiv \alpha \ast \beta
\label{starmult}
\end{eqnarray}
where $\theta^{\mu \nu}$ is a constant antisymmetric tensor.

Let
\begin{equation}
F_\theta = e^{\frac{i}{2} \partial_\mu \otimes \theta^{\mu \nu}
  \partial_\nu} = ``{\rm Twist \; element}'' .
\label{twistelt}
\end{equation}
Then
\begin{equation}
m_\theta (\alpha \otimes \beta) = m_0 [F_\theta \alpha \otimes
\beta] \label{starmult1}
\end{equation}
where $m_0$ is the point-wise multiplication map, also defined by
(\ref{starmult}).

The algebra ${\cal A}_\theta({\mathbb R}^N)$, regarded as a vector
space, is a module for ${\cal A}_0({\mathbb R}^N)$. We can show this
as follows.

For any $\alpha \in {\cal A}_\theta({\mathbb R}^N)$, we can define two
operators $\hat{\alpha}^{L,R}$ acting on ${\cal A}_\theta({\mathbb
  R}^N)$:
\begin{equation}
\hat{\alpha}^L \xi = \alpha * \xi, \quad \hat{\alpha}^R \xi = \xi
* \alpha \quad {\rm for} \quad \xi \in {\cal A}_\theta({\mathbb
R}^N) \ ,
\end{equation}
where $*$ is the GM product defined by Eq.(\ref{starmult}) (or,
equivalently, by Eq.(\ref{starmult1})).  The maps $ \alpha \rightarrow
\hat{\alpha}^{L,R}$ have the properties
\begin{eqnarray}
\hat{\alpha}^L \hat{\beta}^L &=& (\hat{\alpha}\hat{\beta})^L, \\
\hat{\alpha}^R \hat{\beta}^R &=& (\hat{\beta}\hat{\alpha})^R, \label{right}\\
{[}\hat{\alpha}^L, \hat{\beta}^R] &=& 0. \label{LRcommute}
\end{eqnarray}
The reversal of $\hat{\alpha}, \hat{\beta}$ on the right-hand side of
(\ref{right}) means that for position operators,
\begin{equation}
{[}\hat{x}^{\mu L}, \hat{x}^{\nu L}] = i \theta^{\mu \nu} =
-[\hat{x}^{\mu R}, \hat{x}^{\nu R}].
\end{equation}
Hence in view of (\ref{LRcommute}),
\begin{equation}
\hat{x}^{\mu c} = \frac{1}{2} \left( \hat{x}^{\mu L} + \hat{x}^{\mu R}
\right)
\end{equation}
generates a representation of the commutative algebra ${\cal
  A}_0({\mathbb R}^N)$:
\begin{equation}
{[}\hat{x}^{\mu c}, \hat{x}^{\nu c}] = 0.
\end{equation}

Then it is easy to show that for any $\alpha \in {\cal
A}_\theta({\mathbb R}^N)$ one has
\begin{equation}
(\hat{x}^{\mu c} \alpha)(\xi) = \xi^\mu \alpha( \xi)
\end{equation}
and so $\hat{x}^{\mu c}$ generates the commutative algebra ${\cal
  A}_0({\mathbb R}^N)$ acting by point-wise multiplication on ${\cal
  A}_\theta({\mathbb R}^N)$.

This result is implicit in the work of Calmet and coworkers
\cite{calmet1,calmet2}. Let us express ${\rm ad} \hat{x}^\mu$ in
terms of the momentum operator $\hat{p}_\mu = -i \partial_\mu$.
This is easily done using explicit expression for the
star-product, Eq.(\ref{starmult}):
\begin{equation}
{\rm ad} \hat{x}^\mu \alpha = x^\mu * \alpha - \alpha * x^\mu =
i\theta^{\mu\nu}\partial_\nu \alpha = -\theta^{\mu\nu}\hat{p}_\nu\
. \label{adxandp}
\end{equation}

Hence\footnote{If $x^{\mu_0}$ is a commutative coordinate, then
$\theta^{\mu_0 \mu}=0,$ $\forall \mu$, and $\hat{x}^{\mu_0
L}\equiv x^{\mu_0}$ so $\hat{x}^{\mu_0 c}$ is just a usual
commutative coordinate.}
\begin{equation}
\hat{x}^{\mu c} = \hat{x}^{\mu L} - \frac{1}{2} {\rm ad}
\hat{x}^\mu = \hat{x}^{\mu L} + \frac{1}{2} \theta^{\mu \nu}
\hat{p}_\nu.
\end{equation}
This result is the starting point of the work of Calmet et al
\cite{calmet1,calmet2}.

The connected Lorentz group ${\cal L}_+^\uparrow$ acts on functions
  $\alpha \in {\cal A}_\theta({\mathbb R}^N)$ in just the usual way in
  the approach with the coproduct-twist:
\begin{equation}
[U(\Lambda)\alpha](x) = \alpha(\Lambda^{-1} x)
\label{globaltrans}
\end{equation}
for $\Lambda \in {\cal L}_+^\uparrow$ and $U:\Lambda \rightarrow
U(\Lambda)$ its representation on functions. Hence the generators
$M_{\mu \nu}$ of ${\cal L}_+^\uparrow$ have the representatives
\begin{equation}
M_{\mu \nu} = \hat{x}_\mu^c p_\nu - \hat{x}_\nu^c p_\mu, \quad
p_\mu = -i \partial_\mu \label{Ltrans}
\end{equation}
on ${\cal A}_\theta({\mathbb R}^N)$.

The twist is exactly what is required by the coproduct
$\Delta_\theta \equiv F_\theta^{-1} \Delta_0 F_\theta$ \cite{cknt}:
\begin{eqnarray}
\Delta_\theta(M_{\mu \nu}) &=& \Delta_0(M_{\mu \nu}) - \frac{1}{2}
\big[ (p \cdot \theta)_\mu \otimes p_\nu - p_\nu \otimes (p \cdot
  \theta)_\mu - \mu \leftrightarrow \nu \big] \, , \label{twistedcoprod}\\
\Delta_0(M_{\mu \nu}) &=& M_{\mu \nu} \otimes {\bf 1} + {\bf 1}
\otimes M_{\mu \nu} \, .
\end{eqnarray}
Thus
\begin{equation}
m_\theta[\Delta_\theta(M_{\mu \nu}) \alpha \otimes \beta] = M_{\mu
  \nu}(\alpha * \beta).
\end{equation}

\section{A Model for Noncommutative Quantum Electrodynamics}

In \cite{Balachandran:2007kv}, we argued that the group of gauge transformations should be based in the commutative algebra ${\cal A}_0 ({\mathbb R}^N)$. An immediate consequence of this requirement is that the gauge fields $A^\mu$ depend on commutative coordinates $x^c_\mu$ only. In addition, we also required that the quantum covariant derivative $D_\mu$ of quantized matter field $\phi(x)$ transform correctly under twisted (anti-)symmetrization. 

It is an interesting exercise to investigate the consequences of dropping this requirement on the covariant derivative. In other words, what kind of phenomenological signals can be expected if we work with gauge fields that depend only on $x^c$'s but do not insist on twisted covariance for the covariant derivatives? 

Not surprisingly, we find that this leads to a loss of the connection between spin and statistics, which we will demonstrate explicitly in the case of M\"oller scattering, and Compton effect. For example, in the center-of-mass frame, the scattering amplitude ${\cal T}_\theta$ for electrons with
their spin states identical no longer vanishes for $90^o$ scattering angle, violating Pauli principle.

The standard QED in ordinary commutative space is based on the interaction Hamiltonian
\begin{equation}
H_I  = e \int d^3 x \bar{\psi}(x) A^\mu(x) \gamma_\mu \psi(x).
\end{equation}

We will work with its simplest generalization 
\begin{equation}\label{intham}
H_I(t) = e \int d^3 x \bar{\psi}(\hat{x}) * \big( \not \!\!A (\hat{x}^c)
\psi(\hat{x}) \big)
\end{equation}

\subsection{$e^- -e^-$ scattering for general $\theta^{\mu\nu}$}

The Dirac fields are expanded as
\begin{eqnarray}\label{fields}
\psi(\hat x) &=& \int d\mu(k) \sum_s \big[ a^{(s)}(k) u^{(s)}(k)
e^{-ik \cdot \hat x} + b^{\dagger (s)}(k) v^{(s)}(k) e^{i k \cdot \hat
x} \big], \\
\bar{\psi}(\hat x) &=& \int d\mu(k) \sum_s \big[ b^{(s)}(k)
\bar{v}^{(s)}(k) e^{-ik \cdot \hat x} + a^{\dagger (s)}(k)
\bar{u}^{(s)}(k) e^{i k \cdot \hat x} \big]
\end{eqnarray}
while for the gauge field $A_\mu$, we have the expansion
\begin{equation}
\not  \!\! A(\hat x^c) = \int d\mu (k) \sum_r \big[\alpha^{(r)}(k)
\not
  \!\epsilon ^{(r)}(k) e^{-i k \cdot \hat x^c} + \alpha^{(r)\dagger}(k)
  \not \!\bar{\epsilon} ^{(r)}(k) e^{i k \cdot \hat x^c} \big]\ .
\end{equation}
We work in the Lorentz gauge.

The Dirac fields are functions of noncommutative coordinates and its
creation and annihilation operators satisfy twisted (anti) commutation
relations \cite{bmpv}:
\begin{eqnarray}\label{twistcr}
a^{(s_1)}(p_1) a^{(s_2)}(p_2) &=& - e^{i p_1 \wedge p_2}
a^{(s_2)}(p_2)
a^{(s_1)}(p_1) , \\
a^{(s_1)}(p_1) a^{(s_2)\dagger }(p_2) &=& - e^{-i p_1 \wedge p_2}
a^{(s_2)\dagger }(p_2) a^{(s_1)}(p_1) + 2p_{10} \delta (\vec{p}_1 -
\vec{p}_2), \\
a^{(s_1)\dagger }(p_1) a^{(s_2)\dagger }(p_2) &=& - e^{i p_1 \wedge
  p_2} a^{(s_2)\dagger }(p_2) a^{(s_1)\dagger }(p_1).
\end{eqnarray}
Using the map
\begin{equation}
a^{(s)}(p) = c^{(s)}(p) e^{-\frac{i}{2} p \wedge P}, \quad p \wedge P
:= p_\mu \theta^{\mu \nu} P_\nu,
\label{commmap}
\end{equation}
the twisted commutation relations can be realized in terms of the
usual operators $c^{(s)}(p)\equiv
a^{(s)}(p)\mid_{\theta^{\mu\nu}=0}$. Here
\begin{equation}
P_\nu = \int d \mu (k) k_\nu \sum_{s=1,2}(a^{\dagger (s)}(k)
a^{(s)}(k) - b^{\dagger (s)}(k) b^{(s)}(k))
\end{equation}
is the energy-momentum operator of just the electron field in Fock
space.

Similar relations exist for the positron creation and annihilation
operators $b^{(s)\dagger}(k)$ and $b^{(s)}(k)$. \footnote{Let us
comment on the statistics loss under the gauge transformation. One
can easily show that Eq.(\ref{commmap}) is equivalent to the
following map for the field:
$\psi(x)=\psi_c(x)e^{\frac{1}{2}\overleftarrow{\partial}\wedge P}$
where $\psi_c(x)$ is a commutative spinor field. Using covariant
derivative as defined in \cite{Balachandran:2007kv}, one immediately checks
that $D_\mu (\psi_c(x)e^{\frac{1}{2}\overleftarrow{\partial}\wedge
P})\ne (
D_\mu\psi_c(x))e^{\frac{1}{2}\overleftarrow{\partial}\wedge P}$.
Though this is not a very pleasant feature of this formalism and is
partially responsible for the violation of Lorentz invariance
(see the last section), in \cite{Balachandran:2009an}
it is shown that the Lorentz invariance remains only as a
quasi-Hopf algebra even when gauge transformations agree with
statistics.}

The gauge field $A_\mu$ is a function of $\hat{x}_\mu^c$ which
generates the commutative substructure in ${\cal A}_\theta
({\mathbb R}^4)$. Its creation/annihilation operators satisfy the
usual commutation relations:
\begin{equation}
[\alpha^{(r_1)}(k_1), \alpha^{(r_2) \dagger}(k_2)] = - \eta^{r_1 r_2}
2k_{10} \delta (\vec{k}_1 - \vec{k}_2).
\end{equation}

The $S$-matrix for this theory in the interaction representation is
\begin{equation}
T\Big[ \exp -i \int H_I (t) \Big]
\end{equation}
which, as usual, may be expanded in powers of the coupling constant
$e$.

We are interested in $e^- -e^-$ scattering, so the incident and
outgoing state vectors are, respectively
\begin{eqnarray}
|p_1,s_1;p_2,s_2 \rangle &=& a^{(s_1)\dagger} (p_1) a^{(s_2)\dagger}
(p_2) |0\rangle = e^{\frac{i}{2}p_1 \wedge p_2} c^{(s_1)\dagger} (p_1)
c^{(s_2)\dagger} (p_2)|0\rangle, \\
|p'_1,s'_1;p'_2s'_2 \rangle &=& \langle 0 |a^{(s'_1)} (p'_1)
a^{(s'_2)} (p'_2) = e^{-\frac{i}{2}p'_1 \wedge p'_2} \langle 0
|c^{(s'_1)} (p'_1) c^{(s'_2)} (p'_2).
\end{eqnarray}
The first non-trivial contribution to scattering comes from terms
second order in $e$ (or equivalently, first order in the fine
structure constant $\alpha$):
\begin{equation}
{\cal T}_\theta = \frac{(-ie)^2}{2} \int d^4 x_1 d^4 x_2 \Big(
\theta(x_{10}-x_{20}) \bar{\psi}(x_1) * \big( \not \!\!A (x_1^c)
\psi(x)\big) \bar{\psi}(x_2) * \big( \not \!\!A (x_2^c)
\psi(x_2)\big) + x_1 \leftrightarrow x_2 \Big)
\end{equation}
Since positron fields do not contribute to $e^- -e^-$ scattering
at this order, we will ignore them henceforth.

\begin{figure}
\centerline{\epsfig{figure=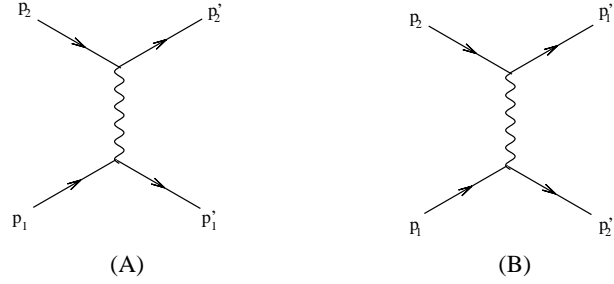,clip=6cm,width=8cm}}
\caption{Feynman diagrams for M\"oller scattering}
\label{fig:moller}
\end{figure}

A long but straightforward calculation then gives us
\begin{eqnarray}
{\cal T}_\theta &=& e^{\frac{i}{2}(p_1 \wedge p_2 - p'_1 \wedge
p'_2)}\left( \frac{\bar{u}^{(s'_1)}(p'_1) \gamma^\mu u^{(s_1)}(p_1)
\bar{u}^{(s'_2)}(p'_2) \gamma_\mu u^{(s_2)}(p_2)}{(p_1 - p'_1)^2}
e^{-\frac{i}{2} (p'_1 \wedge p_1 + p'_2 \wedge p_2)} \right. \nonumber \\
&-& \left.\frac{\bar{u}^{(s'_2)}(p'_2) \gamma^\mu u^{(s_1)}(p_1)
\bar{u}^{(s'_1)}(p'_1) \gamma_\mu u^{(s_2)}(p_2)}{(p_1 - p'_2)^2}
e^{\frac{i}{2} (p'_1 \wedge p_1 + p'_2 \wedge p_2)}
\label{mcmoller1}\right) \\
&=& e^{\frac{i}{2}(p_1 \wedge p_2 - p'_1 \wedge p'_2)} ({\cal T}_1
e^{-i \lambda} - {\cal T}_2 e^{i \lambda}) \label{ncmoller2}.
\end{eqnarray}
Here the first and second terms in correspond to (A) and (B)
respectively of Fig. \ref{fig:moller}.

Notice that we recover the usual answer for M\"oller scattering in the
limit $\theta^{\mu \nu} \rightarrow 0$. Also, there is now a
$\theta^{\mu\nu}$-dependent relative phase between the two terms, and
that will have an observable effect in cross-sections. Secondly, we
anticipate that because of the presence of this relative phase factor,
UV-IR mixing may reappear at higher loops, but this needs to be
checked explicitly.

Let us study the behavior of $\cal T_\theta$ in the center-of-mass
system when the incoming electrons, call them 1,2, have their
spins aligned. We will use the chiral basis. The simplest way to
obtain the electron wave functions is to start with electrons at
rest and boost them along direction $\vec{p}_i = p \hat{p}_i$ and
$\vec{p}_i = -p \hat{p}_i$ for electrons 1 and 2 respectively.
Here the index $i$ stands for initial and, later, $f$ will denote
final. In the rest frame, let the wavefunction of electron 1 be
\begin{equation}
u_{rest}(m,\xi^{(1)}_i) = \left( \begin{array}{c}
                                    \xi_i\\
                    \xi_i
                  \end{array} \right)
\end{equation}
For concreteness, we choose $\xi^{(1)}_i$ to be the spin-up state
along the axis $\hat{p}_i$:
\begin{equation}
\xi^{(1)}_i = D^{\frac{1}{2}}(\hat{p}_i) \left( \begin{array}{c}
                                    1\\
                    0
                  \end{array} \right)
\end{equation}
where $D^{\frac{1}{2}}(\hat{p}_i)$ is the rotation matrix
corresponding to rotation from $z$-axis to direction $\hat p_i$.

Then applying the boost
\begin{equation}
\Lambda^{(1/2)}(\eta) = \exp \left[-\frac{\eta}{2}\left(\begin{array}{cc}
                                   \vec{\sigma}\cdot \hat{p}_i & 0 \\
                   0 & -\vec{\sigma}\cdot \hat{p}_i
                                            \end{array} \right)\right]
\end{equation}
to $u_{rest}(m,\xi^{(1)}_i)$, we get
\begin{equation}
u(p_1, \xi^{(1)}_i) = \left(\begin{array}{c}
                            e^{-\eta/2} \xi^{(1)}_i\\
                e^{\eta/2} \xi^{(1)}_i
                 \end{array}\right)
\end{equation}
where of course $m \sinh \eta = |\vec{p}|$.

Similarly,
\begin{eqnarray}
u(p_2, \xi^{(2)}) &=& \left(\begin{array}{c}
                            e^{-\eta/2} \xi^{(2)} \\
                e^{\eta/2} \xi^{(2)}
                 \end{array}\right) \\
u(p'_1, \xi^{(1)}_f) &=& \left(\begin{array}{c}
                            e^{-\eta/2} \xi^{(1)}_i \\
                e^{\eta/2} \xi^{(1)}_i
                 \end{array}\right) \\
u(p'_2, \xi^{(2)}_f) &=& \left(\begin{array}{c}
                            e^{-\eta/2} \xi^{(2)}_i \\
                e^{\eta/2} \xi^{(2)}_i
                 \end{array}\right)
\end{eqnarray}
where
\begin{equation}
\xi^{(2)}_i = D^{\frac{1}{2}}(\hat{p}_i) \left( \begin{array}{c}
                                    0\\
                    1
                  \end{array} \right)
\end{equation}
Also, in the center-of-mass system, the relative phase $\lambda$ in
(\ref{ncmoller2}) is
\begin{eqnarray}
\lambda &=& \frac{1}{2}(p'_1 \wedge p_1 + p'_2 \wedge p_2)
\theta_{ij}{p'}_1^i p_1^j  \\
&=& \frac{1}{2} \theta_{ij}\epsilon^{ijk} (\hat{p}_f \times
\hat{p}_i)^k (E^2-m^2)\\
&\equiv& \frac{1}{2} m^2 (\vec{T}\cdot \hat{n}) \sin \Theta_M
 \left(\frac{E^2}{m^2}-1\right)
\end{eqnarray}
where $T^i = \theta_{ij} \epsilon^{ijk}$, $\hat{n}$ is the unit
vector normal to the plane spanned by $\hat{p}_i$ and $\hat{p}_f$,
and $\Theta_M$ is the scattering angle. Thus in the center-of-mass
system, all information about noncommutativity is encapsulated in
the scalar product $T \cdot \hat{n}$.

Using the identities
\begin{eqnarray}
|\xi^{(1)\dagger}_f \xi^{(2)}_i|^2 &=& \frac{1}{2} (1-\hat{p}_f \cdot
 \hat{p}_i), \\
\xi^{(1)\dagger}_f \xi^{(1)}_i \xi^{(2)\dagger}_f \xi^{(2)\dagger}_i
 &=& \frac{1}{2} (1+\hat{p}_f \cdot \hat{p}_i),
\end{eqnarray}
it is easy to see that the scattering amplitude for noncommutative
M\"{o}ller scattering (upto an overall numerical factor coming from
normalization of the Dirac spinors) is
\begin{equation}
{\cal T}_\theta = \frac{2 (\hat{p}_f \cdot \hat{p}_i) \cosh 2 \eta
\cos
  \lambda - i \sin \lambda[2(\hat{p}_f \cdot \hat{p}_i)^2 \cosh 2 \eta
  + 1- (\hat{p}_f \cdot \hat{p}_i)^2]}{-2m^2 \sinh^2 \eta [1-
  (\hat{p}_f \cdot \hat{p}_i)^2]}
\end{equation}
${\cal T}_\theta$ is complex in general, and has no real roots. We
can instead look at $|{\cal T}_\theta|^2$. For this, we define
dimensionless quantities $x = E/m$ and  $t= m^2 (\vec{T} \cdot
\hat{n})$. Then
\begin{eqnarray}
|{\cal T}_\theta|^2 &=& \frac{4((2x^2-1)^2 \csc^2 \Theta_M
 +(2x^4-3x^2+1)\cos (t(x^2-1) \sin \Theta_m)-1)\cot^2
 \Theta_M}{4(x^2-1)^2} \nonumber \\
&+& \frac{\sin^2 (\frac{1}{2}t(x^2-1) \sin \Theta_M)}{4(x^2-1)^2}.
\end{eqnarray}

For fixed energy $E$, the minimum of $|{\cal T}_\theta|^2$ (as a
function of $\Theta_M$) is still at $\Theta_M = \pi/2$. This minimum
value is
\begin{equation}
|{\cal T}_\theta|^2_{(\Theta_M = \pi/2)}
=\frac{1-\cos(t(x^2-1))}{8(x^2-1)^2}
\end{equation}

Let us define
\begin{equation}
|{\cal F}|^2 = |{\cal T}_\theta(t, \Theta_M, x)|^2/|{\cal
 T}_\theta(0,\Pi/4, x)|^2
\end{equation}
to rid us of normalization-related ambiguities, and plot $|{\cal
  F}|^2$ as a function of the scattering angle $\Theta_M$.
\begin{figure}
\centerline{\epsfig{figure=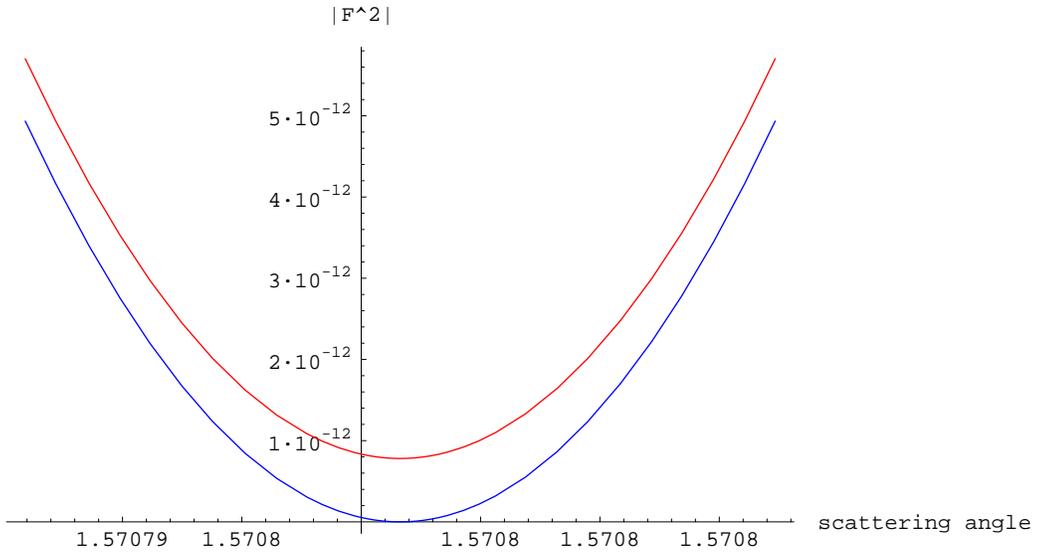,clip=,width=.9\linewidth}}
\caption{$|{\cal F}|^2$ for $t=10^{-5}$ and $x=100$.}
\label{fig:sca1}
\end{figure}

The mod of the squares of the amplitudes are plotted for the
noncommutative and the ordinary cases in Fig \ref{fig:sca1}, where we
see that the noncommutative amplitude does not vanish at
$\Theta_M=\pi/2$.

\begin{figure}
\centerline{\epsfig{figure=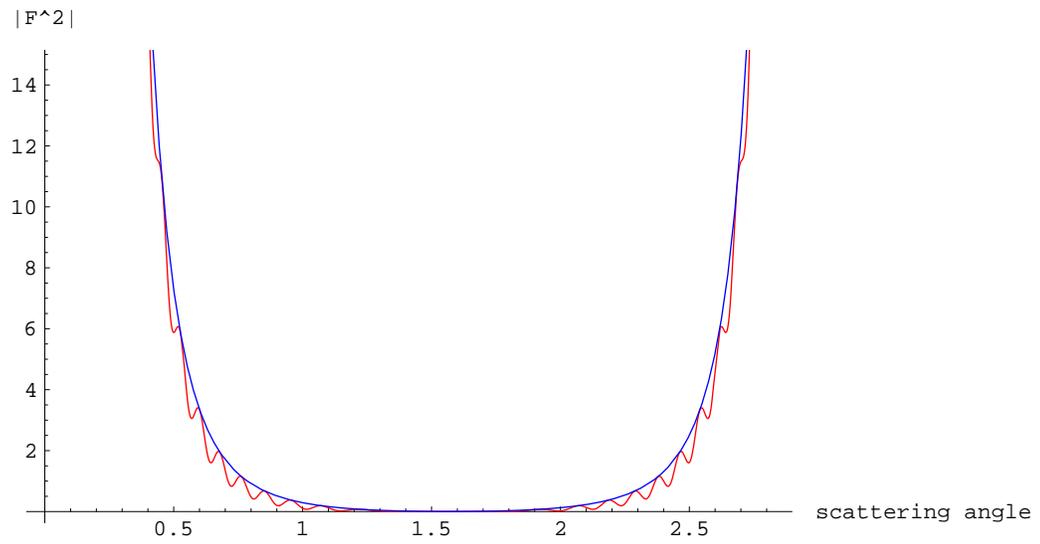,clip=6cm,width=.9\linewidth}}
\caption{$|{\cal F}|^2$ for $t=10^{-2}$ and $x=100$.}
\label{fig:sca2}
\end{figure}

Fig \ref{fig:sca2} shows the same process for larger values of the
scattering angle. We have chosen a much larger value of $t$ to
demonstrate that the noncommutative M\"oller scattering has
characteristic modulations.

\subsection{$e^--e^-$ and Compton scattering when $\theta^{0i}=0$}

In this subsection, we analyze noncommutative QED using a slightly
different approach which can be efficient in the case of just
space-space noncommutativity, i.e. when $\theta^{0i}=0$. In this
case the only Moyal star present in Eq.(\ref{intham}) can be
removed so that the interaction Hamiltonian becomes
\begin{equation}\label{intham1}
H_I(t) = e \int d^3 x \bar{\psi}(x) \not \!\!A (x) \psi(x) \ .
\end{equation}
Though this looks like the interaction Hamiltonian in the
commutative case, there is still the effect of noncommutativity
hidden in the twisted commutation relations (\ref{twistcr}).

Using the map (\ref{commmap}), we can map noncommutative fermionic
fields to the commutative ones,
\begin{eqnarray}\label{fieldmap}
\psi (x)
&=&\psi_0(x)e^{\frac{1}{2}\overleftarrow{\partial}_x\wedge
{P}} \ ,\\
\overline{\psi} (x) &=&\overline{\psi}_0(x) e^{\frac{1}{2}
    \overleftarrow{\partial}_x\wedge
{P}}\ ,\nonumber
\end{eqnarray}
where $\psi_0(x)$ and $\overline{\psi}_0(x)$ are defined as in
(\ref{fields}) for $\theta^{\mu\nu}=0$. We will need one important
property which holds for any two twisted fermionic fields
$a(x)=a_0(x)e^{\frac{1}{2}\overleftarrow{\partial}_x\wedge {P}}$
and $b(x)=b_0(x)e^{\frac{1}{2}\overleftarrow{\partial}_x\wedge
{P}}$ (they can be $\psi$ and/or $\bar\psi$) which is a trivial
consequence of (\ref{fieldmap}):
\begin{equation}\label{inducestar}
a(x)b(x)= \big((a_0\star
b_0)(x)\big)e^{\frac{1}{2}\overleftarrow{\partial}_x\wedge {P}}\ ,
\end{equation}
where $\star:=*_{-\theta}$ is the ``inverse Moyal star''. Using
(\ref{inducestar}) we can write the interaction Hamiltonian as
follows
\begin{equation}
H_I(t) = e \int d^3 x A_\mu (x)\big\{\big( \bar{\psi}_0\star
\gamma^\mu\psi_0(x)\big)e^{\frac{1}{2}\overleftarrow{\partial}_x\wedge
{P}}\big\} \ .
\end{equation}
In the same way, we have for $H_I(t_x)H_I(t_y)$
\begin{eqnarray}
\lefteqn{H_I(t_x)H_I(t_y) = } \nonumber \\
&&e^2 \int d^3 x d^3 y A_\mu (x)A_\nu (y)\big\{\big(
\bar{\psi}_0(x)\star
\gamma^\mu\psi_0(x)\star_{xy}\bar{\psi}_0(y)\star
\gamma^\nu\psi_0(y)\big) e^{\frac{1}{2}(\overleftarrow{\partial}_x
+ \overleftarrow{\partial}_y)\wedge {P}}\big\} \label{hh}
\end{eqnarray}
where $\star_{xy}:=e^{-\frac{i}{2}\overleftarrow{\partial}_x
  \wedge\overrightarrow{\partial}_y}$
(so $\star \equiv \star_{xx}$).

Let us sketch how this approach works in the two cases of
$e^--e^-$ scattering and Compton effect.

\textit{\textbf{A. $e^--e^-$ scattering.}} The first observation we
make is that the factor $e^{\frac{1}{2}(\overleftarrow{\partial}_x
+ \overleftarrow{\partial}_y)\wedge {P}}$ does not contribute to
the result.\footnote{This is because $A_\mu (x)A_\nu (y)$ gives
rise to the photon propagator $G$ which satisfies $(\partial_x +
\partial_y)G(x-y)=0$. So one can extend the action of
$e^{\frac{1}{2}(\overleftarrow{\partial}_x +
\overleftarrow{\partial}_y) \wedge {P}}$ to the whole integrand in
Eq.(\ref{hh}). Integration by parts gives the desired result.} For
the case of $e^--e^-$ scattering the relevant term in
$H_I(t_x)H_I(t_y)$ is hence
\begin{equation}\label{hhee}
e^2 \int d^3 x d^3 y A_\mu (x)A_\nu (y)\big[
\bar{\psi}_0^{(+)}(x)\star
\gamma^\mu\psi_0^{(-)}(x)\star_{xy}\bar{\psi}_0^{(+)}(y)\star
\gamma^\nu\psi_0^{(-)}(y)\big]\ ,
\end{equation}
where $(\pm )$ denote the positive and negative frequency modes.
Using the definition of $\star$, we see that in momentum space, the
overall effect of noncommutativity is the phase
\begin{equation}\label{phase}
e^{-\frac{i}{2}k_1\wedge k_3-\frac{i}{2}q_1\wedge
q_3+\frac{i}{2}(k_1-k_3)\wedge (q_1 - q_3)}\ ,
\end{equation}
where $k_i,q_i$ are integration variables.

Due to momentum conservation at every vertex, one immediately has
$(k_1-k_3)\wedge (q_1 - q_3)=0$.

\begin{figure}
\centerline{\epsfig{figure=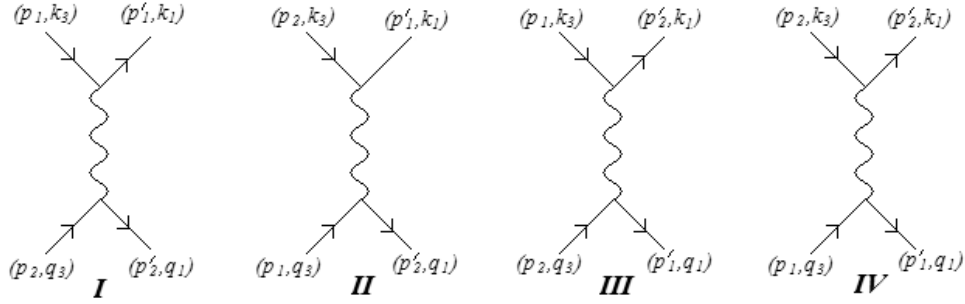,clip=,width=0.9\linewidth}}
\caption{Feynman diagrams relevant for $e-e$ scattering.}
\label{fig:diag1}
\end{figure}

There four different ways to assign external momenta $p_i,p'_i$.
They correspond to four Feynman diagrams (see Fig
\ref{fig:diag1}):
\begin{eqnarray}
(I)& & p_1 = k_3,\ p_2 = q_3,\ p'_1 = k_1\ , p'_2 = q_1 \nonumber
\\
(II)& & p_2 = k_3,\ p_1 = q_3,\ p'_1 = k_1\ , p'_2 = q_1 \nonumber
\\
(III)& & p_1 = k_3,\ p_2 = q_3,\ p'_2 = k_1\ , p'_1 = q_1
\nonumber
\\
(IV)& & p_2 = k_3,\ p_1 = q_3,\ p'_2 = k_1\ , p'_1 = q_1 \nonumber
\end{eqnarray}
In the commutative case, diagrams (I) and (IV) are equal. The same
is true for the diagrams (II) and (III). One can easily see that
this is also the case in the presence of noncommutativity, but now
two different diagrams have a relative phase. Including the
trivial phase factor $e^{\frac{i}{2}(p_1\wedge p_2-p'_1\wedge
p'_2)}$ (this factor comes from the twisted statistics of
\textit{ingoing} and \textit{outgoing} states), we recover the result
(\ref{ncmoller2}).

\begin{figure}
\centerline{\epsfig{figure=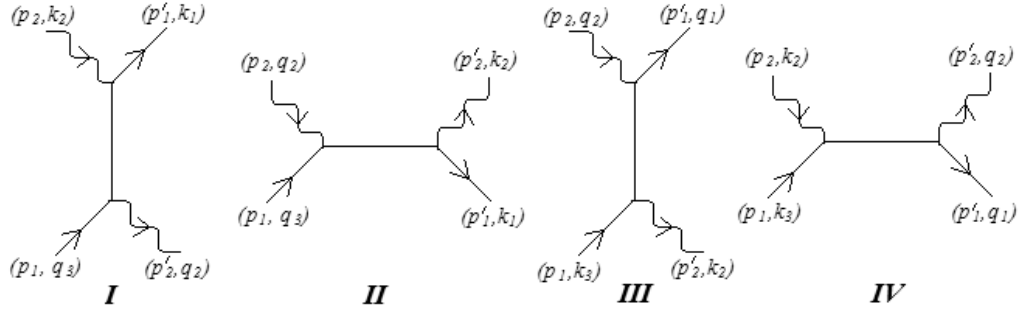,clip=,width=0.9\linewidth}}
\caption{Feynman diagrams relevant for Compton scattering.}
\label{fig:diag2}
\end{figure}

\textit{\textbf{B. Compton Effect.}} The main difference from the
previous case is that we cannot get rid of the factor
$e^{\frac{1}{2}(\overleftarrow{\partial}_x +
\overleftarrow{\partial}_y)\wedge {P}}$ in Eq.(\ref{hh}). This is
due to the fact that now instead of the photon propagator, we have
a spinorial one, which comes from the pairing of fields with twisted
statistics. Nevertheless using the approach developed in
\cite{Pinzul:2005gx}, one can show that the propagator is the same
as in the commutative case and enters all calculations without
star products. In particular this means that all noncommutative
phases will be independent of the momenta of the fields that form
the propagator. Bearing this in mind, one can easily write all
phases. As in the case of $e^--e^-$ scattering, we have four
different possibilities (see Fig \ref{fig:diag2}):
\begin{eqnarray}
(I)& & p_1 = q_3,\ p_2 = k_2,\ p'_1 = k_1\ , p'_2 = q_2 ,\nonumber
\\
(II)& & p_2 = q_2,\ p_1 = q_3,\ p'_1 = k_1\ , p'_2 = k_2 ,\nonumber
\\
(III)& & p_2 = q_2,\ p_1 = k_3,\ p'_2 = k_2\ , p'_1 = q_1,\nonumber
\\
(IV)& & p_1 = k_3,\ p_2 = k_2,\ p'_2 = q_2\ , p'_1 = q_1.\nonumber
\end{eqnarray}
($k_2,q_2$ are the photon momenta. This is why they were absent
for the $e^--e^-$ scattering.) In the commutative case, diagrams
(I) and (III) are equal. The same is true for the diagrams (II)
and (IV). Here the similarities with the case of $e^--e^-$
scattering end. Whereas in that case before the integration over
$k_i,q_i$ the phases were equal to Eq.(\ref{phase}) for all four
diagrams, here the situation is different:
\begin{eqnarray}
\mbox{Phases for }(I)\mbox{ and }(II): & &
e^{-\frac{i}{2}k_1\wedge
q_3}e^{\frac{i}{2}(k_1-q_3)\wedge p_1}\nonumber \\
\mbox{Phases for }(III)\mbox{ and }(IV): & &
e^{-\frac{i}{2}k_3\wedge q_1}e^{\frac{i}{2}(-k_3+q_1)\wedge
p_1}\nonumber
\end{eqnarray}
The second factor in both cases comes when
$e^{\frac{1}{2}(\overleftarrow{\partial}_x +
\overleftarrow{\partial}_y)\wedge {P}}$ picks up the momentum $p_1$ of
the incoming electron. As a result, in the case of Compton scattering,
the effect of the interference between two commutative diagrams is not
present. One rather has that the commutative amplitude
$\mathcal{T}_{com}$ gets multiplied by the overall $\theta$-dependent
factor:
\begin{equation}
\mathcal{T}_{nc} = \frac{1+e^{-i p_1\wedge
p'_1}}{2}\mathcal{T}_{com}\ .
\end{equation}
\begin{figure}
\centerline{\epsfig{figure=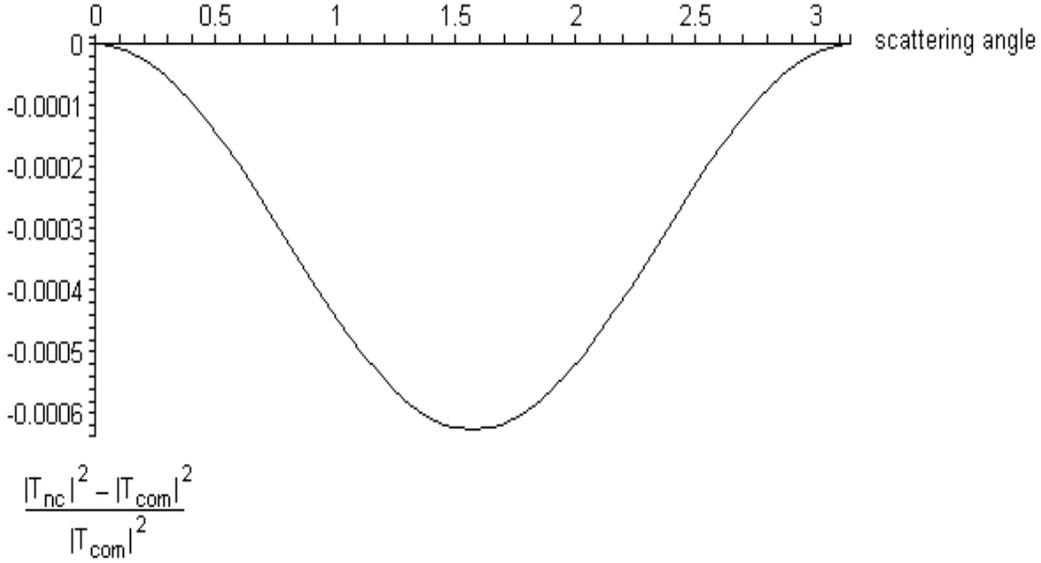,clip=,width=0.9\linewidth}}
\caption{$\frac{|\mathcal T_{nc}|^2- |\mathcal T_{com}|^2}{|\mathcal
    T_{com}|^2}$ as a function of the scattering angle for Compton
    scattering, for $t=10^{-5}$ and $x=100$}
\label{fig:compton}
\end{figure}

Fig. \ref{fig:compton} plots the relative deviation of the scattering
cross-section between the noncommutative and commutative cases for
Compton scattering.

\section{Causality and Lorentz Invariance}

For the purposes of our discussion, causality will have the meaning it
takes in standard local quantum field theories. Thus if $\rho(\xi)$ is
an observable local field $\rho$ like the electric charge density
localized at a spacetime point $\xi$, and $x$ and $y$ are spacelike
separated points ($x \sim y$), then (local) causality states that
\begin{equation}
[\rho(x),\rho(y)]=0.
\label{causality}
\end{equation}
It means that they are simultaneously measurable.

Causal set theory (see for example \cite{causalreview} for a recent
review) uses a sense of causality which differs from
(\ref{causality}). There is also a criticism of the conceptual
foundations of (\ref{causality}) by Sorkin \cite{sorkin}.

Let $H_I$ be the interaction Hamiltonian density in the interaction
representation. The interaction representation $S$-matrix is
\begin{equation}
S = T \exp \left( -i \int d^4 x H_I(x) \right) .
\label{Sformal}
\end{equation}
Bogoliubov and Shirkov \cite{bog} long ago deduced from causality and
relativistic invariance that $H_I$ is a local field:
\begin{equation}
[H_I(x), H_I(y)]=0, x \sim y
\label{localHam}
\end{equation}

Later Weinberg discussed \cite{weinberg1,weinbergbook1} discussed the
fundamental significance of (\ref{localHam}): if (\ref{localHam})
fails, $S$ is not relativistically invariant. He argued as follows.

Let us assume the conventional transformation law for $H_I$ under
Lorentz transformation $\Lambda$:
\begin{equation}
H_I(x) \rightarrow H_I(\Lambda^{-1} x).
\label{LtransH}
\end{equation}
Then we can argue that $S$ is Lorentz invariant provided time-ordering
$T$ does not spoil it.

We can see this as follows. In the absence of time-ordering $T$, the
second order term in $H_I$ in (\ref{Sformal}) is
\begin{equation}
\frac{(-i)^2}{2!} \int d^4x d^4y  H_I(x) H_I(y) = -\frac{1}{2}
\tilde{H}_I(0) \tilde{H}_I(0)
\label{withoutT}
\end{equation}
where $\tilde{H}_I(0)$ is the zero four-momentum component of
$H_I(x)$:
\begin{equation}
\tilde{H}_I(p) = \int \frac{d^4 x}{(2\pi)^4} e^{i p \cdot x} H_I(x)
\end{equation}
We must transform (\ref{withoutT}) according to
(\ref{twistedcoprod}). As zero four-momentum states are both
translation- and Lorentz-invariant, we see that (\ref{withoutT}) is
Lorentz invariant under the twisted coproduct. This argument extends
to all orders in $H_I$ (assuming that long-range (infrared) effects do
not spoil it).

Consider next the second order term $S^{(2)}$ in $S$. It is the leading
term influenced by time-ordering:
\begin{equation}
S^{(2)} = \frac{(-i)^2}{2!} \int d^4x d^4y T(H_I(x) H_I(y)).
\end{equation}
Now
\begin{eqnarray}
T(H_I(x) H_I(y)) &=& \theta(x_0 - y_0) H_I(x) H_I(y) + x
\leftrightarrow y \\
&=&  H_I(x) H_I(y) - \theta(y_0 - x_0)[H_I(x), H_I(y)] \, .
\end{eqnarray}
Under a Lorentz transformation $\Lambda$,
\begin{equation}
 U(\Lambda) T(H_I(x) H_I(y)) U(\Lambda)^{-1} = H_I(\Lambda^{-1} x)
 H_I(\Lambda^{-1} y) - \theta(y_0 - x_0)[H_I(\Lambda^{-1} x),
 H_I(\Lambda^{-1} y)] .
\end{equation}
We exclude the anti-unitary time-reversal from our discussion. Then we see
that $S$ is Lorentz-invariant if
\begin{equation}
\theta(y_0 - x_0)[H_I(\Lambda^{-1} x), H_I(\Lambda^{-1}
  y)]=\theta[\Lambda^{-1}(y-x)_0] [H_I(\Lambda^{-1} x), H_I(\Lambda^{-1}
  y)].
\label{condn1}
\end{equation}
For time- (and light-) like separated $x$ and $y$, $\theta(y_0-x_0)$ is
Lorentz-invariant:
\begin{equation}
\theta(y_0-x_0) = \theta(\Lambda^{-1}(y-x)_0), \quad x \not\sim y \, ,
\end{equation}
and in that case, (\ref{condn1}) is fulfilled. But if $x \sim y$, time
ordering can be reversed by a suitable Lorentz transformation. Hence
Lorentz invariance of $S$ suggests locality:
\begin{equation}
[H_I(x), H_I(y)] = 0 \quad x \sim y \,.
\label{locality}
\end{equation}
[A generalized form of (\ref{locality}) may be enough. See \cite{Balachandran:2007yf}.] Incidentally, 
we cannot say that (\ref{localHam}) [together with
(\ref{LtransH})] is enough for Lorentz invariance. $H_I$ may fulfill
(\ref{localHam}), but $[H_I(x), H_I(y)]$ may contain derivative terms,
such as happens in a charged massive vector meson theory, which spoil
Lorentz invariance \cite{weinberg1}.

The interaction density in the electron-photon system for $\theta^{\mu
  \nu} \neq 0$ is
\begin{equation}
H_I(x) = i e (\bar{\psi}* \gamma^\rho A_\rho \psi)(x) \, .
\end{equation}
For simplicity we consider the case where
\begin{equation}
\theta^{0i} = 0, \quad \theta^{ij} \neq 0
\label{spacelikenc}
\end{equation}
and show that (\ref{localHam}) is violated. Hence $S$ is not
Lorentz-invariant. It can be checked that (\ref{locality}) is violated
if $\theta^{0i} \neq 0$ even if $\theta^{ij}=0$. We can also directly
see from the explicit formula for $e^- -e^-$ scattering amplitude in
Section 8 that it is not Lorentz-invariant if $\theta^{\mu \nu} \neq
0$.

With (\ref{spacelikenc}),
\begin{equation}
S = T \exp\left( -i \int d^3 x H_I(x)\right) = T \exp \left( -i \int
\hat{H}_I(x)\right),\quad
\hat{H}_I(x) = ie (\bar{\psi} \gamma^\rho A_\rho \psi)(x)
\end{equation}
We have used the property of the Moyal product to remove the $*$ from
$H_I$. That is possible although $H_I$ is integrated only over spatial
variables because of (\ref{spacelikenc}). But there is still the
effect of $\theta^{\mu \nu}$ in the oscillator modes of $\psi$ and
$\bar{\psi}$. Let $\psi_0$ be the limit of $\psi$ for $\theta^{\mu
\nu} =0$ and $P_\mu$ be the momentum operator for $\psi$ (which is the
same as for $\psi_0$). Then using (\ref{commmap}),
\begin{eqnarray}
\hat{H}_I(x) &=& J^\lambda (x) A_\lambda(x) \\
J^\lambda (x) &=& J^{(0)\lambda}(x) e^{\frac{1}{2}
  \overleftarrow{\partial}_\mu \theta^{\mu \nu} P_\nu}\, , \\
J^{(0)\lambda}(x) &=& i e \bar{\psi}_0 \gamma^\lambda \psi_0.
\end{eqnarray}
As $A_\lambda$ is not affected by twisting, $[A_\lambda(x),A_\rho(y)]$
is zero for $x \sim y$. The entire effect of $\theta^{\mu \nu}$ is in
$J^\lambda$. But
\begin{equation}
J^\lambda (x) J^\rho (y) \neq J^\rho (y) J^\lambda (x), x \sim y,
\label{nccurrent}
\end{equation}
because of the exponential following $J^{(0) \lambda}$. One can check
(\ref{nccurrent}) by retaining just a pair of distinct momentum modes
in $J^\lambda (x)$ and another such pair in $J^\rho (y)$.

Thus $S$ is not Lorentz invariant.

\bigskip

{\bf Acknowledgments:} It is a pleasure to thank T. R.
Govindarajan for suggesting that we look for zeroes of $e^- -e^-$
scattering. The work of APB, AP and BQ is supported in part by DOE
under grant number DE-FG02-85ER40231.

\end{document}